\def\year{2020}\relax
\documentclass[letterpaper]{article} 
\usepackage{aaai20}  
\usepackage{times}  
\usepackage{helvet} 
\usepackage{courier}  
\usepackage[hyphens]{url}  
\usepackage{graphicx} 
\usepackage{booktabs}
\usepackage{subcaption}
\usepackage{todonotes}
\usepackage{comment}
\usepackage{graphicx}  
\usepackage{color}
\newcommand{\citet}[1]{\citeauthor{#1} \shortcite{#1}}
\newcommand*\samethanks[1][\value{footnote}]{\footnotemark[#1]}

\title{Unsupervised Detection of Sub-events in Large Scale Disasters}

\begin{document}
\setcounter{secnumdepth}{0}  
 \author{\normalsize Chidubem Arachie\thanks{Equal contribution. Research work was done while authors were interning at Dataminr Inc.},\textsuperscript{1}
Manas Gaur\samethanks,\textsuperscript{2}
Sam Anzaroot,\textsuperscript{3}
{\normalsize William Groves,\textsuperscript{3}
Ke Zhang,\textsuperscript{3}
Alejandro Jaimes\textsuperscript{3}
}\\
\textsuperscript{1}{Department of Computer Science, Virginia Tech, Blacksburg, VA, USA achid17@vt.edu}\\
\textsuperscript{2}{Artificial Intelligence Institute, University of South Carolina, Columbia, SC, USA mgaur@email.sc.edu}\\
\textsuperscript{3}{Dataminr Inc., NY, USA}\\
\{sanzaroot, wgroves, kzhang, ajaimes\}@dataminr.com
}
\maketitle
\begin{abstract}

Social media plays a major role during and after major natural disasters (e.g., hurricanes, large-scale fires, etc.), as people ``on the ground'' post useful information on what is actually happening. Given the large amounts of posts, a major challenge is identifying the information that is useful and actionable. Emergency responders are largely interested in finding out what events are taking place so they can properly plan and deploy resources. In this paper we address the problem of automatically identifying important sub-events (within a large-scale emergency ``event'', such as a hurricane). In particular, we present a novel, unsupervised  learning  framework  to  detect sub-events in Tweets for retrospective crisis analysis. We first extract noun-verb pairs and phrases from raw tweets as sub-event candidates. Then, we learn a semantic embedding of extracted noun-verb pairs and phrases, and rank them against a crisis-specific ontology. We filter out noisy and irrelevant information then cluster the noun-verb pairs and phrases so that the top-ranked ones describe the most important sub-events. Through quantitative experiments on two large crisis data sets (Hurricane Harvey and the 2015 Nepal Earthquake), we demonstrate the effectiveness of our approach over the state-of-the-art. Our qualitative evaluation shows better performance compared to our baseline.
\end{abstract}

\section{Introduction}
\label{sec:intro}
Social media has become an essential tool for emergency response during crisis events~\cite{cheng2014event}. As a crisis unfolds, people at the scene of the crisis post critical information about the event on social media in real-time. The data is used by authorities and relief organizations for response planning and relief coordination. However, the massive influx of tweets often prevents humanitarian organizations from being able to make timely judgments. Additionally, the unstructured and informal nature of social media prevents the effective extraction of useful information. Furthermore, the information requirements vary as different stakeholders have different information needs. For example, first emergency responders require fine-grained situational awareness (e.g., shelter needs, first aid, or roads blocked), whereas policymakers require coarse-grained information (e.g., public health awareness, economic breakdowns, or political issues).

In the context of our current study, we define an \textbf{event} as a large scale disaster that causes massive devastation (e.g., a Hurricane). Such large scale emergencies include many important \textbf{sub-events} (e.g., a bridge collapses, power outages, drug shortages, etc.). We detect such events in social media posts (more specifically, on Twitter), and group them into collections called \textbf{sub-event clusters}. These clusters provide a high-level understanding of the crisis and help discover important sub-events. 
For example, during Hurricane Harvey, a Tweet stated: \textit{``power outage in west kingman due to flooding''}. ``Flooding,'' and ``power outage'' are sub-events of Hurricane Harvey.

Our method expands on recent work that models sub-events \textit{exclusively} as noun-verb pairs~\cite{rudra2018identifying}. Nouns are entities, names, or places, while verbs describe actions related to the entity. Noun-verb pairs can represent many, but \textit{not all} sub-events\footnote{Although one could argue that strictly speaking, events can only be described by verbs, in practice, emergency planners and first responders are interested in what's covered by a wider definition that could include topics or themes. For simplicity, we use ``sub-events'' to refer to events and topics or themes of interest.}. Tweets such as \textit{we need to be prepared for infectious diseases that may spread when water recedes \#chennaifloods} and \textit{contaminated water still pose health risks to residents in harvey affected area \#harvey \#texas} describes infectious diseases and contaminated water as highly relevant sub-events but do not conform to the noun-verb pair structure.  To extract these kinds of sub-events, we use a two-word phrase detection model that captures frequent phrases. Our approach combines noun-verb pairs and phrases as a more comprehensive approach for sub-event detection. On large datasets, the number of automatically identified noun-verb pairs and phrases is large and requires pruning. 
Our method identifies the most important sub-events by ranking candidate sub-events using the Management of Crisis (MOAC) ontology\footnote{\url{http://observedchange.com/moac/ns/}} and discarding noun-verb pairs and phrases that do not occur more than once in the dataset. Subsequently, we cluster the top-ranked noun-verb pairs and phrases. 

In summary, our main contributions are as follows: 
\begin{itemize}
    \item We utilize dependency parsing to extract noun-verb pairs from tweets~\cite{rudra2018identifying} and combine them with phrases we extract to form candidate sub-events.
    \item We rank the candidate sub-events by comparing the cosine similarity of their vector representations with vector representations of classes in the MOAC crisis ontology to identify the most important sub-events. 
\end{itemize}
    Previous work has focused exclusively on using noun verb pairs to detect events (or, in other contexts, performing topic modeling). In addition, to the best of our knowledge, we are the first to employ this domain-specific knowledge for ranking of events in the crisis domain. 

We test our approach using tweets from Hurricane Harvey (2017) and Nepal earthquake (2015). We evaluate our method quantitatively (compare to state-of-the-art method for sub-event detection in crisis domain) and qualitatively (by using human annotators through crowdsourcing to determine sub-event cluster quality).

\section{Related Work}
\label{sec:related}
Recent studies have shown that a lot of people rely on social media for information during crisis events \cite{Nazer2017,reuter2017social}. Sometimes it is difficult to filter the correct information given the deluge of chatter on social media. Hence, assessing the right information at the right time from the right sources becomes essential for making life-saving decisions in crises \cite{zade2018situational,chauhan2017providing}. To this end, researchers have extensively studied many aspects of assessing, classifying or analyzing social media content to process them into actionable messages \cite{imran2015processing,Leavitt}.

One aspect of crisis management is identifying sub-events as a significant crisis unfolds \cite{Abhik}. Studies have tried to detect sub-events from tweets using different methods, both supervised and unsupervised \cite{ardalanevent,chen2018encoder,PradhanBOW}.
Some recent supervised methods, introduced the problem of event type detection as a sequence labeling task using a neural sequence model on a news corpus \cite{bekoulis2019sub}. Deep learning techniques using Convolutional Neural Networks and Long Short-Term Memory have been proposed for the task of event/sub-event detection from social media data \cite{nguyen2017robust,Burel2017,pichotta2016learning}. Semi-supervised approaches have also been explored for solving this task, especially concerning crisis events \cite{alam2018graph,alametal2018}.
Though supervised and semi-supervised methods tend to perform well on some tasks, they rely on lexical or syntactic features, which require rigorous human engineering \cite{sha2018jointly}. Further, the efficiency of the supervised method depends on the granularity and quality of annotations. The performance of the model could be reduced depending on the type and ambiguity of the event. In essence, supervised approaches for event detection tend to not generalize well for different crisis scenarios.
Unsupervised sub-event identification methods that have been explored involve identifying topics from tweets by using either Hierarchical Dirichlet Process \cite{SRIJITH2017989}, Self Organizing Maps \cite{Pohl} or Bi-term Topic Modeling \cite{yan2013biterm}. These methods have shown some promise in other natural language processing tasks; however, they are limited to sub-event identification. They identify top frequency words or topics that are not necessarily excellent representatives of sub-events \cite{vieweg2014integrating}. 

``Rapid Automatic Keyphrase Extraction'' (RAKE), is a language-independent tool developed to extract pairs of keywords from structured documents (e.g., news or scientific articles) \cite{rose2010automatic}. RAKE searches for keyword pairs that co-occur at least twice, while maintaining the order, in a document. Other studies, such as \cite{meladianos2015degeneracy} utilize a graph-degeneracy approach to identify subgraph which represents potential sub-events.
Our work is closely related to the recent work by \cite{rudra2018identifying}, which uses a linguistic approach to solve the problem of sub-event identification. The authors extract noun-verb (NV) pairs using dependency parsing from crisis-related tweets. The nouns are entities while the verbs describe the actions performed by the noun e.g., building collapsed, house burning, etc. Subsequently, the NV pairs (or potential sub-events) identified were ranked using Szymkiewicz-Simpson overlap score and a discounting factor to identify infrequent sub-events. The authors refer to their method as Dependency-Parser-based SUB-event detection (DEPSUB). Our approach differs from their work in that we complement the noun-verb pairs with phrases (or sub-event mentions) that users typically use in crisis communications to describe sub-events. These co-occurring words do not occur as noun-verb pairs in most cases.
Also, we propose a different ranking method that takes into account the semantic representation of words to identify the most pressing and useful sub-events. Lastly, we group our sub-events into categories (or sub-event clusters) for better understanding of the disaster scenario.   

\section{Datasets and Processing}
We use Hurricane Harvey (2017) and Nepal Earthquake (2015) as two case studies for our experiments. We use publicly available tweets related to both crises from CrisisNLP.\footnote{\url{https://crisisnlp.qcri.org/}} The resource provides both unlabeled tweet IDs (concerning privacy constraints) and a small labeled tweet corpus for both Hurricane Harvey and Nepal Earthquake. We used the Twitter Hydrator\footnote{\url{https://github.com/DocNow/hydrator}} to extract tweets associated with the tweet IDs. The labeled tweets were marked as either relevant to the crisis events or not relevant to the crisis. A description of the datasets is provided in Table~\ref{tab:datasets}. We combined the unlabeled and labeled data for developing our model. The results of our approach were compared to the baseline, executed over the same labeled tweets. 
We performed an initial preprocessing of the datasets comprising removal of; (1) white spaces, (2) stop words and words with less than three characters, (3) strings with numeric characters, and (4) hashtags (e.g.,  \#harvey, \#hurricane, \#hurricaneharvey).

\begin{table}[t]
\footnotesize
\begin{center}
\begin{tabular}{p{2.5cm}|p{2.5cm}p{2.5cm}}
    \toprule[1.5pt]
      \textbf{Event Type} & \textbf{Hurricane Harvey} & \textbf{Nepal Earthquake} \\ \midrule
      \textbf{Time period} & Aug 27 - Sept 2, 2017 & April 25-30, 2015\\
      \textbf{Unlabeled \/ tweets} & 4.6 Million & 635, 150 \\
      \textbf{Labeled \/ Tweets} & 4,000 & 4,639 \\
      \textbf{Labels} &  Informative/Not-Informative & Informative/Not-Informative \\
     \bottomrule[1.5pt]
\end{tabular}
\end{center}
\caption{Description of the two large scale unlabeled and labeled tweet datasets used in the study.}
\label{tab:datasets}
\end{table}

{\bf Privacy and Ethics Disclosure:} During a disaster, people often share personal information and usernames. We use only public twitter data, and follow standard practices of anonymization during data acquisition and processing by removing names and tweet handles from the text.

\section{Methodology}
\begin{figure*}[t]
  \centering
    \includegraphics[width=50mm,scale=0.5, 
    trim=7.0cm 1.5cm 7.0cm 1.5cm , angle=-90]{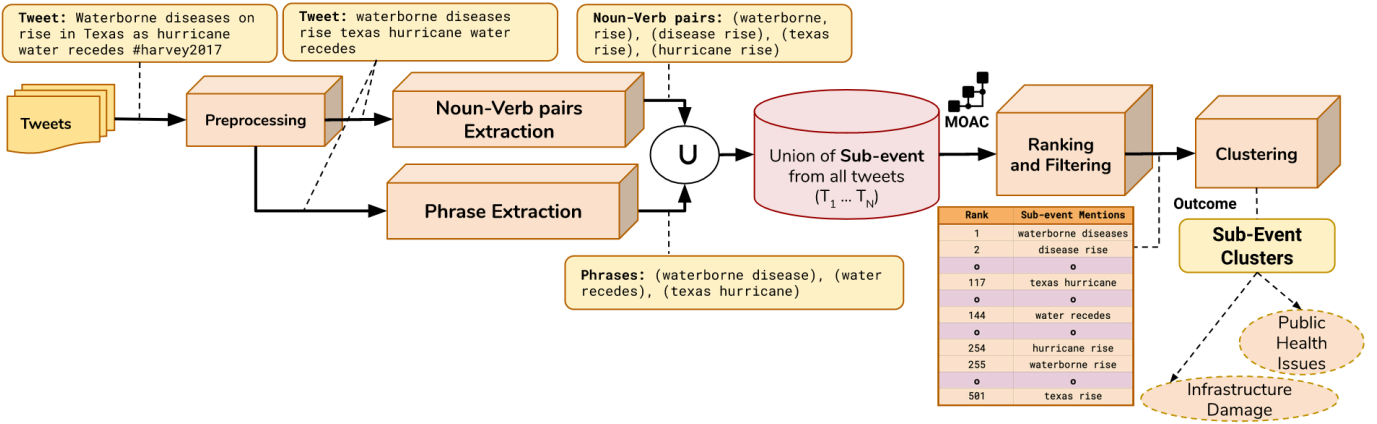}
  \caption{Our framework showing how candidate sub-events are extracted and clustered through an example tweet.}
  \label{fig:architecture}
  \vspace{-1em}
\end{figure*}

In this section, we describe our approach for sub-event identification from tweets and explain how we aggregate the sub-events into sub-event clusters. Our method is divided into three components: (1) extraction of sub-events from text, (2) ranking of candidate sub-events and (3) clustering (see Figure \ref{fig:architecture}). 

\subsection{Extraction of Sub-Events}
We form a single corpus of tweets by appending processed labeled tweets to unlabeled tweets. We utilize the spaCy dependency parser to extract nouns and verbs present in the corpus of tweets~\cite{honnibal2017spacy}. The parser iteratively constructs a dependency tree for each tweet and removes all parent and child nodes that are either nouns or verbs. The number of noun-verb pairs identified from this method is numerous, and a large number of the candidates are not sub-events. To filter out the noisy pairs, we only consider noun-verb pairs that occur more than once in the dataset.  It helps to substantially reduce the number of candidates sub-events without missing out on essential sub-events. 
While noun-verb pairs can identify plenty of sub-events, it misses interesting and implicit sub-events that are not manifested as noun and verb pairs. Hence, we complement them with phrases to capture potential sub-events that occur as co-occurring words. We run a Gensim\footnote{\url{https://radimrehurek.com/gensim/models/phrases.html}} phrase model over the tweet corpus setting the minimum count for co-occurring words as $2$ . Gensim phrase model implements the phrase detection method described in \cite{mikolov2013distributed}.
Consider a sample processed tweet: \textit{waterborne diseases hurricane water recedes}.  The dependency parser identifies ``waterborne recedes'' and ``water recedes'' as noun-verb pairs. The phrase model recognizes ``waterborne diseases'' as a phrase.  Their composition is considered as candidate sub-events. 
\subsection{Ranking of Sub-Events}
Using our method described above, we need to rank the candidate sub-events to identify the most important ones. For example, in the tweet \textit{Waterborne diseases on the rise in Texas as hurricane water recedes \#harvey2017}, ``waterborne diseases'' is a more important sub-event than ``water recedes''. Hence, our ranking method should be able to understand the semantic meaning of the pairs to rank them. To achieve this goal, we use a MOAC ontology containing $62$ terms related to crisis scenarios. We compare our candidate sub-events with terms in the\textit{ MOAC ontology} and \textit{score the max of the cosine similarity between each candidate sub-event and the terms} in the ontology. We need to generate embeddings of our candidate sub-events and terms in the ontology to make the comparisons.  Embeddings provide a numerical vector representation that captures the context of a word in a corpus. Embedding algorithms including Word2Vec \cite{mikolov2013distributed}, GLoVe \cite{pennington2014glove} and FastText \cite{joulin2016fasttext} have proven to be effective for creating rich representations tuned to a specific domain. We trained FastText on $53$ million crisis-related tweets covering Florida Rains (2000 \& 2016), Chennai Floods (2005), Hurricane Sandy (2012), Typhoon Haima (2016), New-Zealand Earthquake (2016), Hurricane Irma (2017), Hurricane Harvey (2017), Houston Floods (2017), and Alaska Earthquake (2018) \cite{gaur2019empathi}. We chose FastText since the method has the ability to generate embeddings for out-of-vocabulary words by leveraging character embeddings. This characteristic of FastText is important due to the noisy nature of social media text, in which there are many misspellings, neologisms, and hashtags. The trained model generated vector representations of tokens in unlabeled tweet datasets. 
The vector representation of a noun-verb pair or phrase, was generated through the summation of individual word vectors. Further, we normalize the word vectors, as our downstream task of ranking and clustering requires computation of cosine similarity between noun-verb pairs or phrases with concepts in MOAC ontology. 
Our ranking approach gives a higher score to candidate sub-events that are semantically related to crisis terms in the MOAC ontology. 

\subsection{Sub-Event Clusters}
Aggregation of candidate sub-events is critical for rapid situational awareness. It will address the use cases of first responders (e.g., navy, firefighters, local emergency manager) and humanitarian organizations. 
We cluster our filtered list of candidate sub-events to enable us to label a cluster as belonging to a type/category (see Figure \ref{fig:architecture}). Our clusters should be diverse and should group related sub-events such that each cluster will represent a cohesive category of sub-event. We investigated various clustering approaches (e.g., DBSCAN, OPTICS, K-Means, and Gaussian Mixture Model) and found spectral clustering gave the most stable clusters for our task~\cite{ng2002spectral}. The manifolds created in spectral clustering involve the creation of the similarity matrix using cosine similarity as a distance measure between the candidate sub-events. Our resulting clusters are topically diverse and evenly distributed as we will show in the qualitative evaluation.

\section{Experiments and Evaluation}
\label{sec:experiments}
We analyze the performance of our approach on two disaster events: Hurricane Harvey and the Nepal Earthquake. 
The success of a sub-event and categorization scheme depends on (1) how accurately it can identify underlying sub-events in the data, and (2) how practical the categories are in describing the event types in the dataset. In this regard, we provide quantitative and qualitative evaluation for our task where they apply.

\noindent\textbf{Baseline Approaches:} We compare our method to a recent state-of-the-art approach for sub-event identification described by~\cite{rudra2018identifying}. The methodology, DEPSUB, outperforms several baseline methods in their evaluation.
DEPSUB uses only noun-verb pairs for sub-event identification. Furthermore, it employs a different ranking scheme: a product of Szymkiewicz-Simpson overlap score of the sub-events and a discounting factor to reduce the count of infrequent sub-events. 

We also compare our method to a variant of the baseline that uses only noun-verb pairs but employs our ranking methodology (MOAC+NV). The difference between our method and this approach illustrates how phrases alone contributes in detecting comprehensive sub-events.

For homogeneity, in comparison, we followed similar pre-processing steps in the baseline study. However, we utilized the spaCy dependency parser as opposed to Twitter POS tagger for the extraction of Noun-Verb pairs. We considered speed and accuracy to be practical with spaCy compared to Twitter POS tagger~\cite{dutt2018savitr}. 

\subsection{Quantitative Evaluation}
For our quantitative evaluation, we compare our approach with the baselines by measuring how good the top-ranked sub-events are at retrieving informative tweets from the annotated datasets. In particular, given the ranked list of sub-events, we pick the top-$k$ sub-events represented by NV-pairs and phrases and check their occurrence in labeled tweets. A true positive is an annotated informative tweet that includes at least one of the top-$k$ sub-events; otherwise, it is a false negative. We measure the precision and recall at different $k$. 
We define precision as the ratio of the number of informative tweets identified to the total number of tweets identified. Our recall is the ratio of the number of identified informative tweets to the total number of informative tweets in the dataset. We report F1-scores at the different $k$ and the ROC curve. 
Through these two metrics, we evaluate the effectiveness of our approach in retrieving informative tweets over the baseline methods. 

\begin{figure*}[htb]
\centering
\begin{subfigure}{0.5\textwidth}
    \centering
    \includegraphics[width=0.9\textwidth]
    {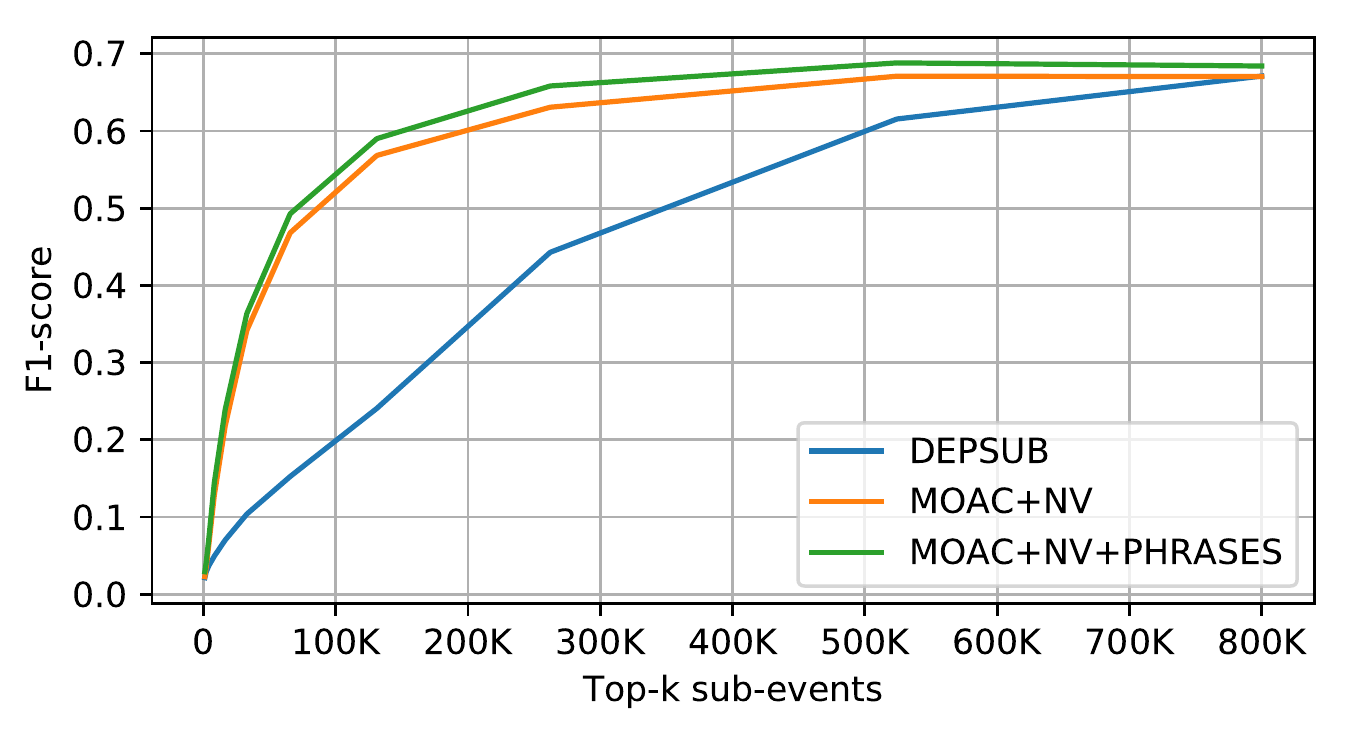}
    \caption{F1-score over varying sub-events thresholds}
    \label{fig:f1harvey}
\end{subfigure}%
\begin{subfigure}{0.5\textwidth}
    \centering
    \includegraphics[width=0.9\textwidth]
    {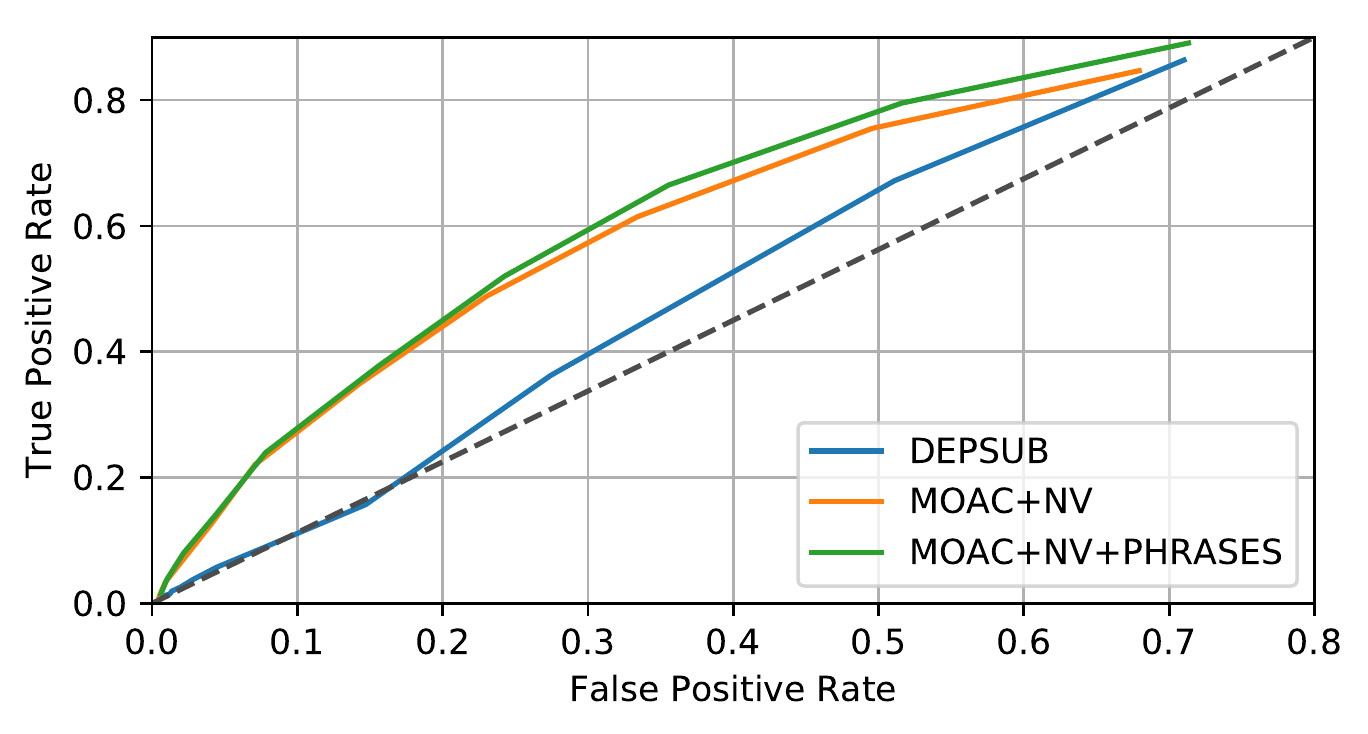}
    \caption{ROC curve}
    \label{fig:rocharvey}
\end{subfigure}
\caption{Assessing the relevance of candidate subevents in identifying informative tweets in labeled Hurricane harvey dataset.  The sub-events were not filtered based of the noun-verb pairs}
\label{fig:harvey-results}
\end{figure*}

\begin{figure}[htb]
    \centering
    \includegraphics[width=0.45\textwidth]
    {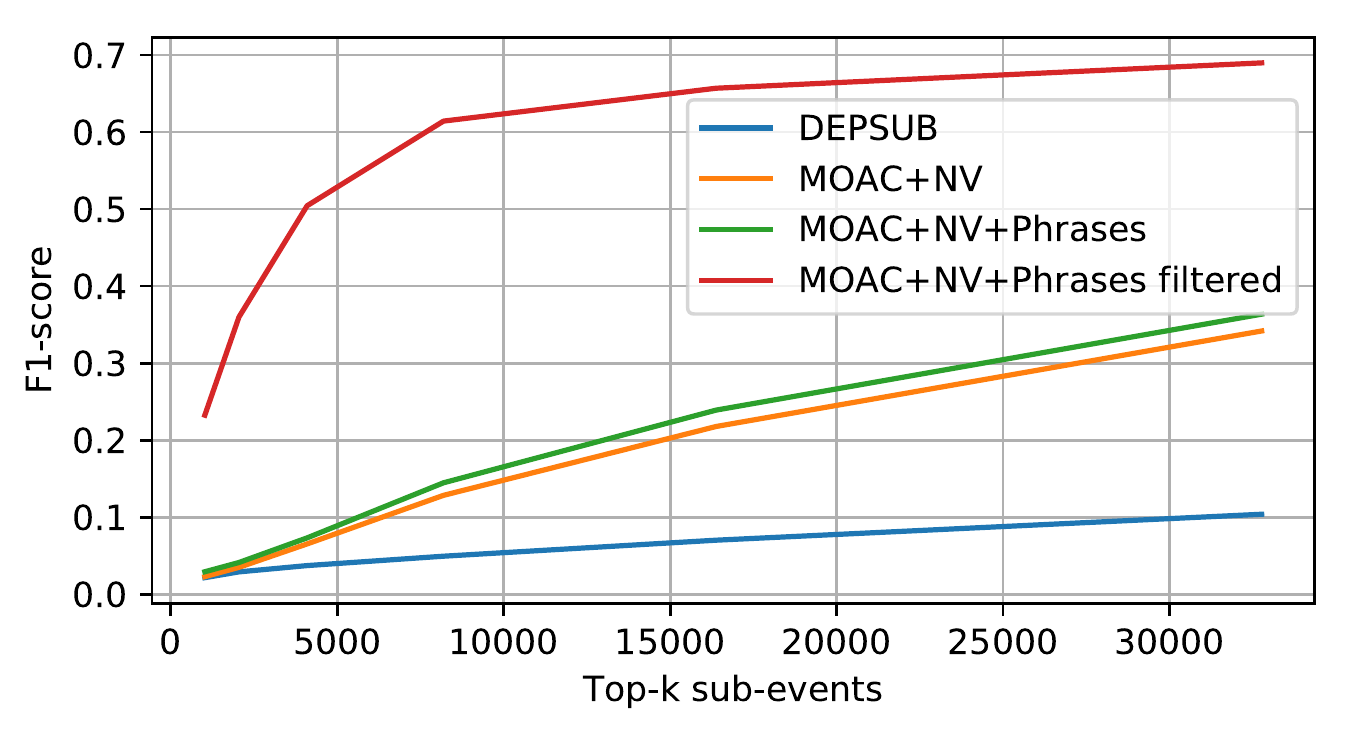}
    \caption{Variation in F1-score on increasing the number of candidate sub-events to extract informative tweets from annotated Hurricane Harvey dataset}
    \label{fig:harvey-results-filtered}
\end{figure}

\noindent\textbf{Hurricane Harvey}: ~~ We used $795,461$ distinct unlabeled tweets from the hydrated $4.6$ million tweets together with $4,000$ ($3,027$ informative and $973$ un-informative) labeled tweets to train the methods. The number of unique noun-verb pairs identified was $769,670$, while the number of phrases totaled $27,122$. Hence, the baseline method (DEPSUB) had $769,670$ candidate sub-events while our approach (without-filtering) has $796,792$. Firstly, we show the performance of our approach compared to the baseline method without our filtering approach. Then, we show the performance of our approach with filtering applied compared to without filtering and to the baseline.

We observe in Figure~\ref{fig:harvey-results} that our approach outperforms the baseline (DEPSUB) in F1-score over a various number of top-ranked sub-events. We also observe the that we obtain marginal gains by adding phrases to noun-verb pairs for sub-event detection. More so, the curve illustrates that our ranking methodology identifies the most important sub-events compared to the ranking used in DEPSUB.
Considering Figure~\ref{fig:f1harvey}, we observe that with the top 250,000 ranked sub-events we achieve optimal results concerning precision and recall of retrieved informative tweets. Additionally, we see in Figure \ref{fig:rocharvey} plot that our method performs well compared to the baseline that slightly performs better than random.

\noindent\textbf{Hurricane Harvey filtered}: We apply our filtering procedure that considers only noun-verb pairs that occur more than once in the tweet corpus. Doing this, we reduce the noun-verb pairs to $3,187$ and the total number of sub-events to $30,309$. This constitutes a $99.6\%$ reduction in the number of noun-verb pairs considered as sub-events and $96.2\%$ reduction in the total number of sub-events. We see from the results of  Figure \ref{fig:harvey-results-filtered} that our filtering approach outperforms the non-filtered methods and the baseline in terms of F1-score. Putting these results in perspective, we have used substantially fewer candidate sub-events to achieve results on the dataset.  Our approach has effectively filtered out non-sub-events from the candidate sub-events.

\noindent\textbf{Nepal Earthquake}: ~~ To confirm the generalizability of our approach, we confirm our results on a different crisis event dataset. We used $635,150$ distinct unlabeled tweets from the hydrated tweets together with $3,479$ ($1,636$ informative and $1,843$ un-informative) labeled tweets to train the methods. The number of unique noun-verb pairs identified was $577,914$, while the number of phrases totaled $36,980$. In this regards, the DEPSUB method had $577,914$ candidate sub-events while our approach (without-filtering) has $614,894$ potential sub-events. As with the previous experiment, we first show the performance of our approach compared to the baseline method without our filtering approach. Then, we show the performance of our approach with filtering applied compared to without filtering and to the baseline.

\begin{figure*}[htb]
\centering
\begin{subfigure}{0.5\textwidth}
    \centering
    \includegraphics[width=0.95\textwidth]
    {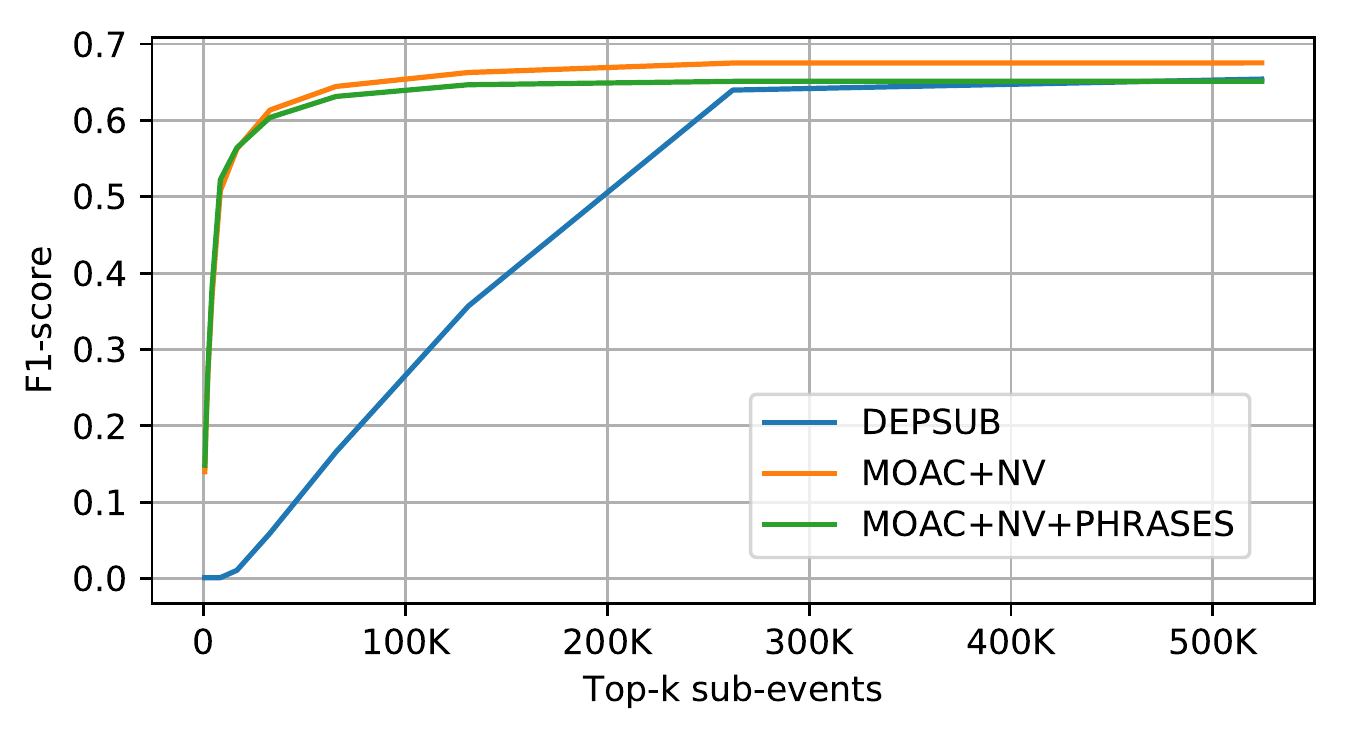}
    \caption{F1-score over varying sub-events thresholds}
    \label{fig:f1nepal}
\end{subfigure}%
\begin{subfigure}{0.5\textwidth}
    \centering
    \includegraphics[width=0.95\textwidth]
    {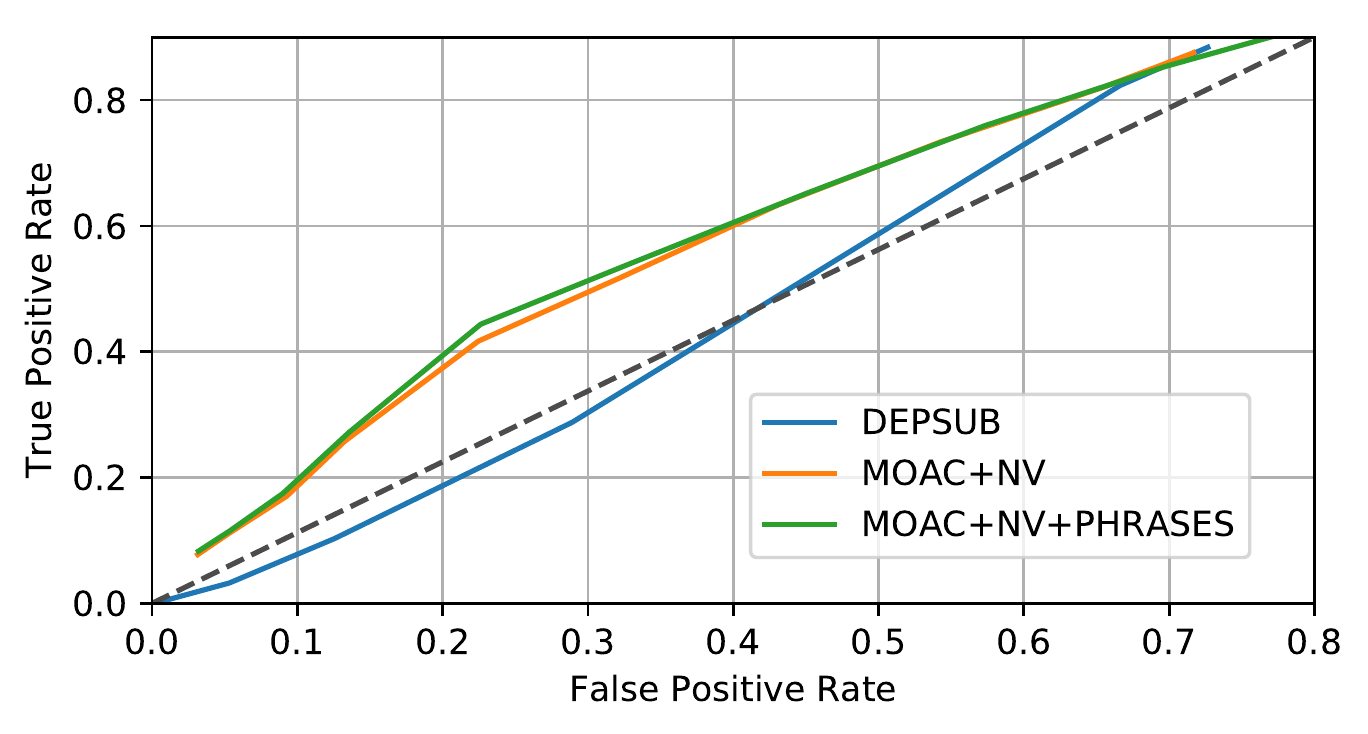}
    \caption{ ROC curve}
    \label{fig:rocnepal}
\end{subfigure}
\caption{Assessing the relevance of candidate sub-events in identifying informative tweets in labeled Nepal Earthquake dataset.  The sub-events were not filtered based of the noun-verb pairs}
\label{fig:nepal-results}
\end{figure*}

We observe in Figure \ref{fig:nepal-results} that our approach outperforms the baseline in F1-score with a fewer number of sub-events than DEPSUB. Additionally, we see in Figure \ref{fig:rocnepal} plot that our method performs better than the baseline method, which performed worse than random for most of the threshold values. It was because the informative tweets retrieved using DEPSUB were \textit{negatively correlated} with the actual result (in other words identified more uninformative tweets).
Furthermore, the utilization of domain-specific crisis embedding model and MOAC ontology enriched our ranking process by up-voting sub-events that are relevant in crisis scenarios. The contribution of phrases alone in this dataset is not as prominent as in the previous experiment but it does help in Figure \ref{fig:rocnepal}.

\begin{figure}[htb]
    \centering
    \includegraphics[width=0.45\textwidth] 
    {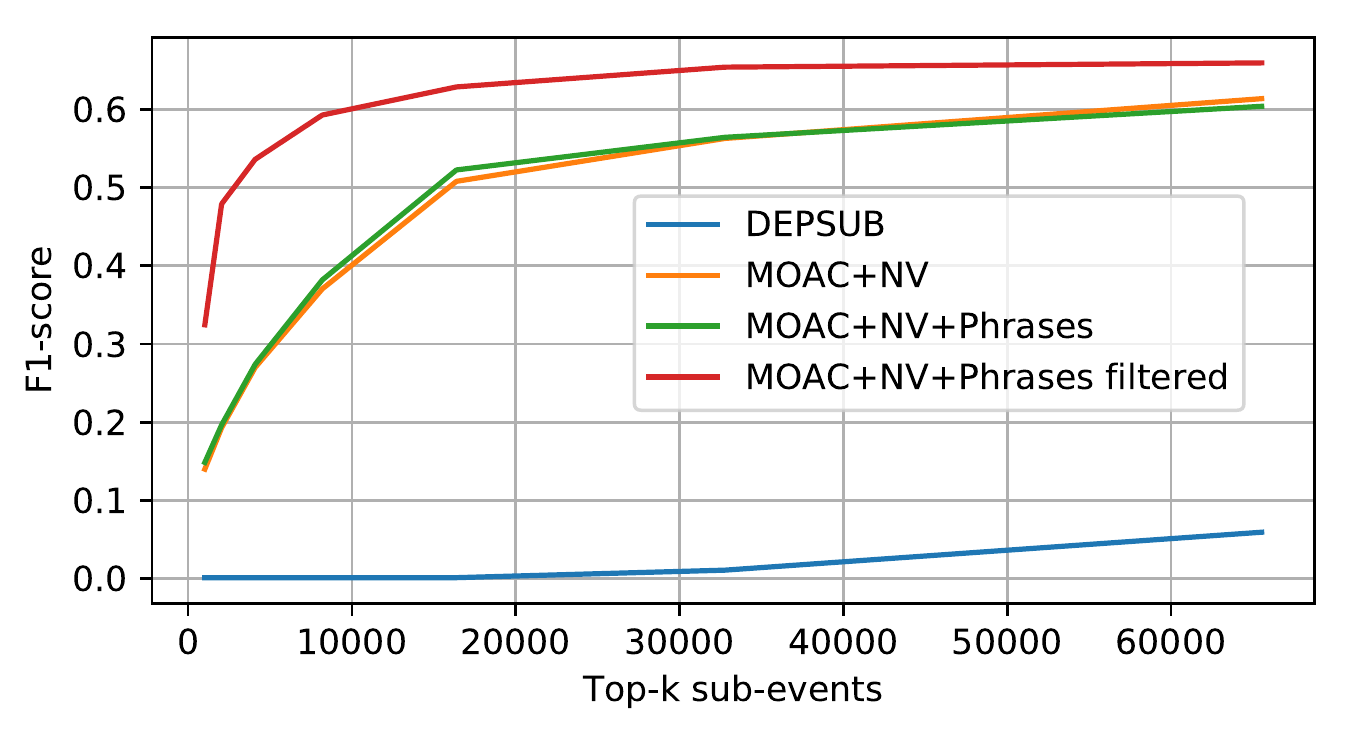}
    \caption{Variation in F1-score on increasing the number of candidate sub-events to extract informative tweets from annotated Nepal Earthquake dataset}
    \label{fig:nepal-results-filtered}
\end{figure}

\noindent\textbf{Nepal Earthquake filtered}: Similar to the first experiment, we apply our filtering procedure that considers only noun-verb pairs that occur more than once in the tweet corpus. Doing this, we reduce the number of noun-verb pairs to $19,229$ and the total number of sub-events to $55,571$. This shows a $96.7\%$ reduction in the number of noun-verb pairs considered as sub-events and a $90.7\%$ total reduction in the number of sub-events. We see from the results of Figure \ref{fig:nepal-results-filtered} that our filtering approach significantly outperforms the non-filtered method and the baseline in terms of F1-score. This result confirms that our approach generalizes over crisis events and substantially reduces noise from the candidate sub-events identified with all noun-verb pairs. 

\subsection{Qualitative Evaluation}
Beyond our quantitative analysis, we also evaluate our methods in terms of quality. We conduct two qualitative evaluations 1) for our sub-events and 2) for the categories in our clusters.

\begin{table}[bt]
\begin{tabular}{p{4cm}p{3.5cm}}
    \toprule[1.5pt]
      \textbf{MOAC-NV+Phrase-filtered} & \textbf{DEPSUB Baseline} \\ \midrule
       feeding centers & foxnews flooding  \\
       road blocked & victims buzzfeed  \\
       shortage fuel  & flotus donated \\
       price gouging  & redcross serving \\
       hundreds trapped  & spca need  \\
       shelter supplies  & mullins flooding  \\
       drug shortage & coldwell impacted  \\
       infectious disease  & sentedcruz impacted  \\
       medical equipment  & peoples lost  \\
       water contamination  & hurr impacted \\
     \bottomrule[1.5pt]
\end{tabular}
\caption{List of top ranked sub-events in Hurricane Harvey dataset}
\label{tab:harveyCSE}
\end{table}

\begin{table}[bt]
\begin{tabular}{p{4cm}p{3.5cm}}
    \toprule[1.5pt]
      \textbf{MOAC-NV+Phrase-filtered} & \textbf{DEPSUB Baseline} \\ \midrule
      internet access  & jeetpur tell  \\
      persons missing & people livez \\
      public health & country redefined \\
      power outage & waves clifton \\
      shelter needs & machineries started \\
      water hygiene & parliament subsidized \\gtfc
      human remains & ayurveda words \\
      riot cops & pepoles lost  \\
      thugs looting & tsunamy trying \\
      reported deaths & chen missing\\
     \bottomrule[1.5pt]
\end{tabular}
\caption{List of top ranked sub-events in Nepal Earthquake dataset}
\label{tab:nepalCSE}
\end{table}

\subsubsection{Ranked Sub-Events}

\begin{figure}[htb]
    \centering
    \includegraphics[width=0.10\textwidth, 
    trim=8.0cm 4.0cm 8.0cm 3.5cm, angle=-90]{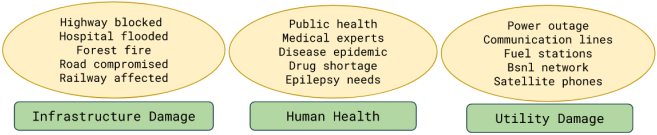}
    \caption{Sample sub-event (oval shape) and sub-event clusters (rectangle shape) in our experiments}
    \label{fig:sampleclusters}
\end{figure}

We posit that a good sub-event identification method should be able to identify important and diverse sub-events.
Tables~\ref{tab:harveyCSE} and~\ref{tab:nepalCSE} show the top ranked sub-events using our MOAC ranking and filtering approach compared to the top ranked sub-events using DEPSUB. We observe that our approach (MOAC-NV+Phrases-filtered) extract important and diverse sub-events compared to DEPSUB. Though the baseline yielded some relevant sub-events (``machineries started'', ``parliament subsidized'', ``flotus donated'', ``redcross serving'') in its top ranks, the other sub-events do not accurately represent incidents that occurred during the crisis events.  Thus, our top sub-events can inform first responders of the most pressing needs during a crisis scenario (see Figure \ref{fig:sampleclusters}).

\begin{table}[ht]
\footnotesize
\begin{center}
\begin{tabular}{p{1.1cm}|p{6.5cm}}
    \toprule[1.5pt]
Label 1 & Emergency Response (e.g. search and rescue, volunteering, donation) \\
Label 2 & Property Damage (e.g. damage, loss) \\
Label 3 & Public Health (e.g. pollution, hospital) \\
Label 4 & Affected Individuals (e.g. injured/missing/found) \\
Label 5 & Security / Public Safety (e.g. violence, theft) \\
Label 6 & Infrastructure and Utility (e.g. electricity, road infrastructure)\\
Label 7 & Politics / Entertainment\\
     \bottomrule[1.5pt]
\end{tabular}
\end{center}
\caption{A collection of crisis-related sub-event type labels derived from the MOAC crisis ontology for the MTurk assessment.}
\label{tab:mturklabels}
\end{table}

\subsection{Human Evaluation of Cluster Quality}

\begin{table}[ht]
\footnotesize
\begin{center}
\begin{tabular}{p{2cm}|p{5.5cm}}
    \toprule[1.5pt]
Treatment A & average number of labels in the collection: $2.00$ \\
Treatment B & average number of labels in the collection: $2.11$ \\
Students' T-test p-value & $0.0205$ \\
Statistical Significance & Yes with p-value $<$ 0.05\\
     \bottomrule[1.5pt]
\end{tabular}
\end{center}
\caption{Student's T-test to show statistical significance of our approach over random baseline.}
\label{tab:mturksig}
\end{table}

\begin{table}[ht]
\footnotesize
\begin{center}
\begin{tabular}{p{1.4cm}|p{2.5cm}|p{2.5cm} }
    \toprule[1.5pt]\
    & Proposed method & Random \\
Labels & (Treatment A) & (Treatment B) \\ \midrule
1  &  $0.31$ & $0.36$ \\
2  &  $0.15$ & $0.11$ \\
3  &  $0.10$ & $0.096$ \\
4  &  $0.12$ & $0.15$ \\
5  & $0.043$ & $0.051$ \\
6  &  $0.050$ & $0.049$ \\
7  &  $0.21$ & $0.18$ \\
     \bottomrule[1.5pt]
\end{tabular}
\end{center}
\caption{Human annotators' label distribution by treatment (where $0.1 = 10\%$).}
\label{tab:mturkdist}
\end{table}

As illustrated in Figure \ref{fig:architecture}, the terminal component of our approach involves clustering the ranked sub-events to identify categories that summarize the sub-events.  Using spectral clustering, we cluster the sub-events generated using our filtering approach. We generated 40 and 50 clusters for Hurricane Harvey and Nepal Earthquake respectively. 
To get a sense of the quality of clusters, we point at an inherent property of spectral clustering: \textbf{Homogeneity} \cite{xu2016effective}. 
Unfortunately, it is difficult to characterize the quality of the clusters with respect to this property, however we describe a crowd-sourced qualitative evaluation using human annotators.
We used the Amazon Mechanical Turk platform to determine that the sub-events within a sub-event cluster are more homogeneous compared to a random baseline.  
Each sub-event cluster is randomly sampled for a set of social media posts, and these posts are presented to a human evaluator as a cohesive collection. The evaluator is asked to provide up to three \textit{sub-event type} labels (Table~\ref{tab:mturklabels}) that best describe the collection. To provide a baseline for comparison, collections of random tweets are provided as an alternative for collection construction. The sampling methodology is summarized as follows:
\begin{itemize}
    \item \textbf{Treatment A:} 100 collections of tweets (5 tweets in each collection) that all belong to the same sub-event cluster using the methodology proposed in this paper.
    \item \textbf{Treatment B:} 100 collections of randomly sampled tweets (5 tweets in each collection) from the entire set of tweets in the Harvey dataset.
\end{itemize}

The annotations provided by the human evaluators show that the sub-event clustering methodology proposed in this paper generates collections of tweets that are significantly more cohesive than a random collection. Results in Table~\ref{tab:mturksig} shows that human annotators provide \textit{fewer} topic labels more often when labeling tweet collections from the sub-event cluster output than when labeling randomly sampled collections of tweets. 

Table~\ref{tab:mturkdist} shows the distribution of labels selected by human annotators for the different methods. Our proposed method (treatment A) is different from the randomly sampled collection case (treatment B). The distributions are observed to have a statistically significant difference using the chi-square test with a p-value of $0.00176$.

\section{Discussion}
\label{sec:discussion}
In our work, we have described the challenges of actionable information delivery through a retrospective study of Hurricane Harvey and Nepal Earthquake using Twitter data. Our method and findings show the positive social impact artificial intelligence can have on society. Through our platform, policymakers and humanitarian organizations can analyze disaster information better for planning and better decision making. 
Although we have summarized the particular information need of our platform for policymakers and first responders, we state how other stakeholders can be assisted in disaster scenarios:

\emph{News Agencies}: suffer from issues concerning incomplete information and time-sensitive matters, causing sparse assemblage of a newsworthy story \cite{mele2017linking}. Our approach can provide a complete situational overview of an event to assist journalists with information dissemination when preparing newsworthy articles. 

\emph{Medical Experts} are alerted during or post-crisis phases when there's a public health issue such as disease outbreak, water contamination, or drug shortages. Hence, early identification of such events through our platform can bolster early intervention.

\section{Conclusion}
A straightforward approach to sub-event identification has been described using publicly available large scale crisis data from Twitter. 
It is composed of three stages requiring, extracting candidate sub-events, ranking and categorization of sub-events to provide better situational information. Through our experiments, we establish that our approach outperforms the current state-of-the-art in sub-event identification from social media data. 
Our framework generalizes to different large scale events within the crisis domain. We have shown this by testing our methodology on two crisis events of different types: a hurricane and an earthquake. These crisis events have different dynamics and can have dissimilar sub-events. The MOAC ontology is general enough to adequately capture the differences. To go beyond the crisis domain, a user of our framework can substitute MOAC with a different ontology that is more tailored to their problem domain. For example, if a user is trying to detect cyberbullying on social media data, they can use an ontology/lexicon that is more aligned with harassment or discrimination for their ranking.

We believe there is much scope for future work in this area as social media is becoming a critical channel for communication during large-scale world events. In the future, we plan to address the real-time nature of the problem by developing an on-line version of our method. It will accommodate novel sub-events that emerge on social media.

\bibliography{References.bib}
\bibliographystyle{aaai}

\end{document}